\documentclass[12pt,showpacs,preprintnumbers,amsmath,amssymb]{revtex4}

\usepackage{epsf}
\usepackage{graphicx}  
\usepackage{dcolumn}   
\usepackage{bm}        

\newcommand{\be}{\begin{equation}}
\newcommand{\en}{\end{equation}}

\begin{document}

\preprint{ITP-CAS/3-2004, \space arxiv:hep-th/0403234}

\title{Inflation on a Warped Dvali-Gabadadze-Porrati Brane}
\author{Rong-Gen Cai\footnote{E-mail: cairg@itp.ac.cn} and
Hongsheng Zhang\footnote{E-mail: zhanghs@itp.ac.cn}}
\affiliation{Institute of Theoretical Physics, Chinese Academy of
Sciences, P.O. Box 2735, Beijing 100080, China}
\date{ \today}

\begin{abstract}
We discuss an inflation model, in which the inflation is driven by
a single scalar field with exponential potential on a warped DGP
brane. In contrast to the power law inflation in standard model,
we find that the inflationary phase can exit spontaneously without
any mechanism. The running of the index of scalar perturbation
spectrum can take an enough large value to match the observation
data, while other parameters are in a reasonable region.

\end{abstract}

\pacs{ 98.80.Cq, 04.50.+h }

\maketitle

\section{Introduction}

Astronomical observations in resent years, especially by Wilkinson
Microwave Anisotropy Probe (WMAP) \cite{wmap}, lead to a high
precision era of cosmology. All the results imply there exists an
accelerating moment (inflationary phase) in the very early
universe. The angle power spectrum of cosmological microwave
background (CMB) provides some evidences that the universe is
almost spatially flat and that large scale structure is formed
from a primordial spectrum of adiabatic, normal and nearly
Harrison-Zeldovich (scale invariant) density perturbations. This
can be explained by the simplest model of
inflation~\cite{liddlelyth}.

Despite the great success of the big bang standard model of
cosmology together with inflation, there are still several serious
problems in our present scenario of cosmology.  The cosmological
constant (dark energy), unexpected low power spectrum at large
scales, egregious running (to the common inflationary models) of
power spectrum index are distinctive ones. The latter two might
relate to the very high energy physics. Although many approaches
have been made in the literature, fairly speaking, these problems
still stick on.

 In view of achievements and shortcomings of inflationary scenario
 in Einstein gravity, it is necessary to further deepen our
 understanding of the inflationary scenario from a theoretical
 perspective. In
nonperturbative string/M theories there are topological solitons,
called branes,  in $10$ or $11$ dimensional spacetime.  In the
Horava-Witten model \cite{horava}, gauge fields of the standard
model are confined on two 9-branes located at the end points of an
$S^1/Z_2$ orbifold, i.e., a circle folded on itself across a
diameter. Inspired by the Horava-Witten model, the idea that our
universe is a 3-brane embedded in a higher dimensional spacetime
has received a great deal of attention in recent
years~\cite{braneworld}. In this brane world scenario, the
standard model particles are confined on the 3-brane, while the
gravitation can propagate in the whole space. In this picture, the
Friedmann equation of the brane universe gets
modified~\cite{branecos}, compared to the one for the standard
model. The chaotic inflation model on the RSII brane has been
studied in~\cite{chaotic}.  The inflation model in the brane world
scenario with a Gauss-Bonnet term in the bulk has also been
discussed recently by Lidsey and Nunes~\cite{LN}.

 In the RSII model~\cite{RSII},  a brane with positive tension is
 embedded in five dimensional anti de Sitter space. Due to the
 warped effect of bulk geometry, general relativity on the brane
 can be recovered in low energies, while
 gravity on the brane is five dimensional in high energy limit.
 Compared to the RSII model, the brane world model
 proposed by Dvali,
 Gabadadze and Porrati (DGP)~\cite{dgpmodel} is also very interesting.
 In the DGP model, the bulk is a flat Minkowski spacetime, but a reduced
 gravity term appears on the brane without tension.
 In this model, gravity appears
4-dimensional at short distances but is altered at distance large
compared to some freely adjustable crossover scale $r_0$ through
the slow evaporation of the graviton off our 4-dimensional brane
world universe into an unseen, yet large, fifth dimension. The
late-time acceleration is driven by the manifestation of the
excruciatingly slow leakage of gravity off our four-dimensional
world into an extra dimension. This offers an alternative
explanation for the current acceleration of the universe.

In the DGP model, the gravitational behaviors on the brane are
commanded by the competition between the 5-dimensional curvature
scalar ${}^{(5)}R$ in the bulk and the 4-dimensional curvature
scalar $R$ on the brane. At short distances the 4-dimensional
curvature scalar $R$ dominates and ensures that gravity looks
4-dimensional. At large distances the 5-dimensional curvature
scalar ${}^{(5)}R$ takes over and gravity spreads into extra
dimension. As a result, Newton-like force law becomes
5-dimensional one. Thus, gravity begins weaker at cosmic
distances. So it is natural that such a dramatic modification
affects the expansion velocity of the Universe. In fact the DGP
model has been applied to cosmology immediately after DGP putting
forward their model \cite{dgpcosmology}.

In this paper, we will study an inflation model in a brane world
scenario, which combines the RSII model and DGP model.  That is,
an induced curvature term will also appear on the brane in the
RSII model. We call the brane as warped DGP brane. In this  model,
inflation of the universe is driven by a single scalar field with
exponential potential on the brane. In contrast to the power law
inflation in standard model, we find that the inflationary phase
exits spontaneously. The running of the index of the scalar
perturbation spectrum can take negative values, which content the
high precision observations by WMAP. At the same time other
parameters are in a reasonable region.

The organization of this paper is as follows. In the next section,
we present the Friedmann equation on the warped DGP brane. In
Secs.~III and IV the inflation model is introduced. We analyze the
parameter space allowed by the observation data in Sec. V. The
paper ends in Sec.~VI with conclusions and discussions.

\section{The Model and Friedmann equation}

We start from the action of the generalized DGP model given in
\cite{eff}
 \be
 \label{eq1}
 S=S_{\rm bulk}+S_{\rm brane},
 \en
where
 \be
 \label{eq2}
  S_{\rm bulk} =\int_{\cal M} d^5X \sqrt{-{}^{(5)}g}
\left[ {1 \over 2 \kappa_5^2} {}^{(5)}R + {}^{(5)}L_{\rm m}
\right],
 \en
and
 \be
 \label{eq3}
 S_{\rm brane}=\int_{M} d^4 x\sqrt{-g} \left[
{1\over\kappa_5^2} K^\pm + L_{\rm brane}(g_{\alpha\beta},\psi)
\right].
 \en
Here $\kappa_5^2$ is the  5-dimensional gravitational constant,
${}^{(5)}R$ and ${}^{(5)}L_{\rm m}$ are the 5-dimensional
curvature scalar and the matter Lagrangian in the bulk,
respectively. $x^\mu ~(\mu=0,1,2,3)$ are the induced 4-dimensional
coordinates on the brane, $K^\pm$ is the trace of extrinsic
curvature on either side of the brane and $L_{\rm
brane}(g_{\alpha\beta},\psi)$ is the effective 4-dimensional
Lagrangian, which is given by a generic functional of the brane
metric $g_{\alpha\beta}$ and matter fields $\psi$ on the brane.

Consider the brane Lagrangian consisting of the following terms
\begin{eqnarray}
\label{eq4}
 L_{\rm brane}=  {\mu^2 \over 2} R -  \lambda + L_{\rm
m},
\end{eqnarray}
where $\mu$ is a parameter with dimension of $[mass]$, $R$ denotes
the curvature scalar on the brane, $\lambda$ is the tension of the
brane, and $L_{\rm m}$ stands for the Lagrangian of other matters
on the brane.
 We assume that there is only a cosmological constant ${}^{(5)}\Lambda$ in
 the bulk.  Therefore the action (\ref{eq1}) describes a generalized DGP model
 or a generalized RSII model, since it goes to the DGP model if
 $\lambda=0$ and ${}^{(5)}\Lambda=0$, or to the RSII model if $\mu =0$.

From the field equations induced on the brane given in \cite{eff},
one can define an effective 4-dimensional cosmological constant on
the brane
 \be
 \label{eq5}
\Lambda={1\over 2} ({}^{(5)}\Lambda+{1\over
6}\kappa_5^4\lambda^2),
 \en
 which is the same as that in the RSII model.
Considering a Friedmann-Robertson-Walker (FRW) metric on the
brane, the dynamical equation of the brane, namely the Friedmann
equation,  is found to be~\cite{eff},
 \be
 \label{eq6}
 H^2+{k\over a^2}={1\over 3\mu^2}\Bigl[\,\rho+\rho_0\bigl(1 +
\epsilon{\cal A}(\rho, a)\bigr)\,\Bigr],
 \en
 where as usual, $k$ is the constant curvature of the maximal symmetric
 space of the FRW metric and $\epsilon$ denotes either $+1$ or $-1$. ${\cal A}$ is
defined by

\be
 {\cal A}=\left[{\cal A}_0^2+{2\eta\over
\rho_0}\left(\rho-\mu^2 {{\cal E}_0\over a^4}
\right)\right]^{1\over 2},
 \en
where
 \be
 {\cal A}_0=\sqrt{1-2\eta{\mu^2\Lambda\over \rho_0}},\ \
\eta={6m_5^6\over \rho_0\mu^2} ~~~(0<\eta\leq 1),
 \en
\be \rho_0=m_\lambda^4+6{m_5^6\over \mu^2}.
 \en
Note that here there are three mass scales, $m_4= \mu$,
$m_\lambda=\lambda^{1/4}$ and $m_5=\kappa_5^{-2/3}$. Further, due
to the appearance of $\epsilon$, there are two branches in this
model, as in the original DGP model. ${\cal E}_0$ is a constant
related to dark radiation~\cite{branecos}. Since we are interested
in the inflation dynamics of the model, as usual, we neglect the
curvature term and dark radiation term in what follows. Then the
Friedmann equation (\ref{eq6}) can be rewritten as
 \be
 \label{eq10}
 H^2=\frac{1}{3\mu^2}\left[\rho+\rho_0+\epsilon \rho_0 ({\cal A}_0^2
 +\frac{2\eta\rho}{\rho_0})^{1/2}\right].
 \en

For the DGP model,  one has $\eta =1$ and ${\cal A}_0=1$. In the
high energy limit $\rho /\rho_0 \gg 1$, the Friedmann equation
reduces to
\begin{equation}
\label{eq11} H^2 \approx \frac{1}{3\mu^2}\left( \rho +\epsilon
\sqrt{2\rho \rho_0}\right).
\end{equation}
This describes a 4-dimensional gravity with a minor correction,
which implies that the parameter $\mu$ must have an energy scale
as the Planck scale: $\mu \sim 10 ^{18} Gev$ in the DGP model. On
the other hand, in the low energy limit $\rho/\rho_0 \ll 1$, the
Friedmann equation (\ref{eq10}) becomes
\begin{equation}
\label{eq12}
 H^2 \approx \frac{1}{3 \mu^2}\left( (1+\epsilon)\rho
 +(1+\epsilon)\rho_0
 -\frac{\epsilon}{4}\frac{\rho^2}{\rho_0}\right).
 \end{equation}
 When $\epsilon =1$, the equation describes a 4-dimensional
 gravity and the term $\rho_0$ gives an effective cosmological constant.
  To be consistent with the current observation, $\rho_0$
 must be in the order of the current critical energy density of
 the universe, namely, $\rho_0 \sim (10^{-3}ev)^4$. When $\epsilon
 =-1$, however, the equation tells us that the gravity on the
 brane is 5-dimensional. This requires that at least $\rho_0 \sim
 (10^{-3}ev)^4$, once again.

 For the warped DGP model with $\lambda \ne 0$ and $^5\Lambda \ne 0$,
 a remarkable point is that the two conditions $\mu \sim
 10^{18}Gev$ and $\rho_0 \sim (10^{-3}ev)^4$ are not necessary.
 To see this, let us consider the limit of $\mu \to 0$ of the
 equation (\ref{eq10}). For simplicity, as in the RSII model, we
 set $\Lambda=0$ in (\ref{eq5}). In this case, we have ${\cal
 A}_0=1$ and $\eta \sim 1-{\cal O}(\mu^2m_{\lambda}^4/m_5^6)$.
 In the ultra high energy limit where
 $ \rho \gg \rho_0 \gg m_{\lambda}^4$, the Friedmann equation
 (\ref{eq10}) is
 \begin{equation}
 \label{eq13}
H^2 \approx \frac{1}{3\mu^2}\left( \rho +\epsilon \sqrt{2\rho
\rho_0}\right).
\end{equation}
This describes a four dimensional gravity on the brane. In the
intermediate energy region where $\rho \ll \rho_0 $ but $\rho \gg
m_{\lambda}^4$, for the branch with $\epsilon =-1$, the Friedamnn
equation changes to
\begin{equation}
\label{eq14}
 H^2  \approx \frac{m_{\lambda}^4}{18m_5^6}\left( \rho
+\frac{\rho^2}{2m_{\lambda}^4} -\frac{\mu^2
m_{\lambda}^4}{6m_5^6}\rho -\frac{\mu^2}{4m_5^6}\rho^2\right).
\end{equation}
This is just the Friedmann equation in the RSII model with some
corrections proportional to $\mu^2$. When $\mu^2=0$, the Friedmann
equation in the RSII model is restored. Finally in the low energy
limit where $\rho \ll \rho_0$ and $\rho \ll m_{\lambda}^4$, we
know from the above equation that the four dimensional general
relativity with the Planck energy $m_p^2 = 6 m_5^6/m_{\lambda}^4$
is recovered on the brane.

In this paper, we do not restrict the small $\mu$ limit, instead
we take the $\mu$ in the almost same order as the Planck scale, as
in the original DGP model. Then in the high energy limit with
$\rho \gg \rho_0$, we have the Friedamnn equation
\begin{equation}
\label{eq15}
H^2 \approx \frac{1}{3\mu^2}\left( \rho +\epsilon
\sqrt{2\rho \rho_0}\right),
\end{equation}
while in the low energy limit with $\rho \ll \rho_0$, for the
branch with $\epsilon =-1$, the Friedamnn equation is
\begin{equation}
\label{eq16} H^2 = \frac{1}{3\mu^2_p}\left (\rho + {\cal O}\left
(\frac{\rho}{\rho_0}\right )^2\right),
\end{equation}
where $\mu^2_p= \mu^2/(1-\eta)$ is the effective four dimensional
Planck scale. As a result, due to the appearance of the tension
$\lambda$, One needs not take $\rho_0$ a very low energy scale
$\rho_0 \sim (10^{-3}ev)^4$ as it is in the original DGP model,
instead one can see from (\ref{eq15}) and (\ref{eq16}) that in the
warped DGP model, $\rho_0$ can be taken at least $\rho_0 \sim (1
Mev)^4$, the BBN energy scale. Note that here the four dimensional
Planck energy scales change from the high energy limit
(\ref{eq15}) to the low energy limit (\ref{eq16}).

In the following discussions, we will not take the equation
(\ref{eq15}) in the high energy limit or the equation (\ref{eq16})
in the low energy limit as our starting point, instead we will
start from the complete equation (\ref{eq10}).   This equation
(\ref{eq10}) shows some new interesting features. A term
proportional to $\rho^{1/2}$, in contrast to $\rho^2$ in the RSII
model, appears. In particular,  a noticeable point is that this
term can be negative. We may expect that  the inflationary
behavior on the brane will be significantly different from that in
the standard model or RSII model. It turns out it is true.

\section{inflationary phase}

As a simple inflation model, we assume that the inflation is
driven by a single scalar field $\phi$ with a potential $V(\phi)$
on the brane. As usual,  the scale field is minimally coupled to
the gravity field. The action of the scalar field then can be
written down as
 \be S_m=\int
d^4x\sqrt{-g} (\frac{1}{2}\partial_\mu \phi
\partial^\mu \phi+V(\phi)).
\en Varying the action yields the equation of motion of the scalar
field
 \be
 \label{eq18}
\ddot{\phi}+3H\dot{\phi}+\frac{dV}{d\phi}=0.
 \en
 The slow roll parameters defined by
 \begin{eqnarray}
&& f\triangleq -\frac{\dot{H}}{H^2},\nonumber \\
 && \alpha\triangleq
\frac{d^2V}{d\phi^2}/(3H^2), \nonumber \\
&& \xi^2\triangleq\frac{d^3V}{d\phi^3}\frac{dV}{d\phi}/(3H^2)^2,
\label{slp}
\end{eqnarray}
can be expressed as
 \be
  f=\frac{\mu^2}{2} \frac{{(\frac{dV}{d\phi})}^2}{V^2}
  \left( \frac{1+\epsilon \eta({\cal
  A}_0^2+\frac{2\eta\rho}{\rho_0})^{1/2}} { \left\{1+\frac{\rho_0}{\rho}
  [1+\epsilon({\cal
  A}_0^2+\frac{2\eta\rho}{\rho_0})^{1/2}]\right\}^2} \right)
 \label{slf}
  \en
 in the slow roll approximation, $(\dot{\phi})^2 \ll V$ and
$\ddot{\phi} \ll |3H \dot{\phi}|$. The other two parameters can
also be calculated similarly.  The term in bracket is the
correction to the standard model. It is easy to see that this term
loosens the condition for inflation when $\epsilon=1$ and tightens
the condition when $\epsilon=-1$, in contrast to the RSII model
where the correction term  always loosens the inflation condition.
The number of e-folds $N\triangleq \ln\frac{a_{end}}{a_0}$, can be
expressed by
 \be
 \label{eq21}
 N=-\int_{\phi_i}^{\phi_{end}} 3H^2  \frac{d\phi}{dV}
d\phi,
 \en
 in the slow roll approximation, where $\phi_{end}$ is the value of
$\phi$ when the universe exits from inflationary phase and
$\phi_i$ denotes the value of $\phi$ when the universe scale
observed today crosses the Hubble horizon during inflation.

Next we investigate the scalar curvature perturbation of the
metric. The perturbations generated quantum mechanically from a
single  field during inflation are adiabatic. The curvature
perturbation on a uniform density hypersurface is conserved on
large scales as a result of energy--momentum conservation on the
brane \cite{wands}. So we expect that as in the RSII model, the
scalar curvature perturbation amplitude of a given mode when
reentering the Hubble radius is still given by
$A_S^2=H^4/(25\pi^2\dot{\phi}^2)$ \cite{liddlelyth} (see also
\cite{LN}). Substituting (\ref{eq18}) with the slow roll
approximation, we obtain
  \be
  \label{eq22}
 A_S^2= \left. \frac{1}{25 \mu^6\pi^2}\frac{9H^6}{(\frac{dV}{d\phi})^2} \right |_{k=aH},
  \en
 The COBE normalization gives us $A_S^2=4\times 10^{-10}$
\cite{cobe}. With the definition of scalar spectrum index
$$n_S=1+d\ln A_S^2/d\ln k,$$  we have
\be
n_S=1-6f+2\alpha,
 \label{nse}
 \en
and the running of the scalar spectrum index
\be \frac{dn_S}{d\ln
k}=16f\alpha-24f^2-2\xi^2.
 \label{rune}
 \en

\section{Exponential potential}

In inflation models, the exponential potential is an important
example which can be solved exactly in the standard model. In
addition, we know that such exponential potentials of scalar
fields occur naturally in some fundamental theories such as
string/M theories.  On the other hand,  we find that the inflation
phase in the warped DGP model even with an exponential potential
on the brane can exit naturally.  It is therefore quite
interesting to investigate such a potential in the warped DGP
model, and to compare its predictions with observational data.

 We write down the potential as
  \be
  \label{eq25}
  V=\widetilde{V}
e^{-\sqrt{2/p}\frac{\phi}{\mu}},
 \en
where $\widetilde{V}$ and $p$  are two constants.
 Calculating the slow roll
parameters,  we have
 \begin{eqnarray}
 &&
 \label{eq26}
 f=\frac{1}{p}
\frac{1+\epsilon \eta({\cal A}_0^2+2\eta x)^{-1/2}}
{\left\{1+\frac{1}{x} [1+\epsilon({\cal A}_0^2+2\eta
x)^{1/2}]\right\}^2},
 \\
&&
  \alpha=\frac{2}{p}\frac{x}{x+1+\epsilon({\cal A}_0^2+2\eta
x)^{1/2}},
 \\
&&
  \xi^2=\alpha^2,
\end{eqnarray}
where $x=\frac{V}{\rho_0}$. We know that when $f=1$, the inflation
ends. If the universe can enter the inflationary phase, i.e.,
$p>2$, it can not exit in the branch of $\epsilon=1$. But we find
that the $f$ in (\ref{eq26}) can increase to one from a value
which is less than one in the branch of $\epsilon =-1$. It implies
that the inflationary phase can exit naturally without any other
mechanism in the branch of $\epsilon =-1$. So in what follows, we
only consider the branch of $\epsilon=-1$.
 However, the equation $f=1$, namely,
\be \label{eq29}
 \frac{1}{p} \frac{1- \eta({\cal A}_0^2+2\eta x)^{-1/2}}
{\left\{1+\frac{1}{x} [1-({\cal A}_0^2+2\eta
x)^{1/2}]\right\}^2}=1
 \en
  is a quintic equation,  whose roots can not be
written down in finite algebraic way. Thus we will give some
numerical results in parameter space. To show that the equation
(\ref{eq29}) could indeed hold, let us first consider a special
case, namely the original DGP model where $\eta={\cal A}_0=1$.  In
this case, we have
 \be
 \lim \limits_{x\to 0}f\to \infty,
 \en
and
 \be \lim \limits_{x\to \infty} f= \frac{1}{p}.
 \en
Because $f$ is a continuous function of $x$, there must exist an
$x\in (0,\infty)$, which satisfies $f=1$ for $p>1$.  So indeed in
the original DGP model the inflation phase can exit spontaneously.
But we find that this happens around $x\sim 1$.  Since $\rho_0
\sim (10^{-3}ev)^4$ in the original DGP model, the inflation would
last to such a low energy scale. This is obviously unreasonable.
In the warped DGP model, however, we find that this difficulty can
be avoided.

Integrating Eq.~(\ref{eq21}), we obtain
\be \label{eq32}
  N=-\left. \frac{p}{2}\left(\ln x-x^{-1}+(Wx^{-1}+
  \frac{2\eta}{{\cal A}_0}\tanh^{-1}(\frac{W}{{\cal A}_0}))\right)\right|_{x_{i}}^{x_{end}},
  \en
 where $W=({\cal A}_0^2 +2\eta x)^{1/2}$, $x_i$ stands for the value of $x$ when the
 cosmic scale observed today crosses the Hubble horizon during inflation. Denote
 \be
  h(x)=-\left(\ln x-x^{-1}+(Wx^{-1}+
  \frac{2\eta}{{\cal A}_0}\tanh^{-1}(\frac{W}{{\cal A}_0}))\right),
    \en
  Eq. (\ref{eq32}) can be expressed as
  \be
  \label{eq34}
  h(x_{end})=h(x_{i})+\frac{2N}{p}.
  \en
Next from the scalar curvature perturbation amplitude
Eq.~(\ref{eq22}), we can get a constraint among $V, ~\rho_0$ and
$p$ \be
 \left. \frac{p}{150\pi^2 V^2\mu^4}\left(V+\rho_0+\epsilon\rho_0
 ({\cal A}_0^2 +\frac{2\eta V}{\rho_0})^{1/2}\right)^3\right|_{k=aH}
 =4\times 10^{-10}.
 \label{cobe}
 \en
Substituting (\ref{slf}) into Eqs.~(\ref{nse}) and (\ref{rune}),
we arrive at
 \be
 n_S=1-\frac{2}{p} \frac{2 {\cal A}_0^2+\eta x+(-2+x)W} {W(1+x-W)},
 \en
\be
 \frac{dn_S}{d\ln k}=\frac{8}{p^2} \frac{x^2 \left\{-{\cal A}_0^4
 +{\cal A}_0^2[-1+2x-2(-1+x)W]
 +\eta x[-2+x(4+\eta-2W)] \right\}} {W^2(1+x-W)^4} ,
 \en
 respectively. Note that when $x\to \infty$,
the results reduce to those of the standard model, as we expected.

\section{Analysis of the parameter space}

For the sake of demonstration, we give an concrete example to show
some interesting properties of this inflationary model on the
warped DGP brane. Because the equation (\ref{eq34}) of the number
of e-folds is very complicated,  we add an additional assumption
that we take $\eta=0.99$, although our method can be used for any
value of $\eta$.  As we mentioned above, the inflation phase can
exit naturally in this model. For example, numerical analysis
shows that the equation $f=1$ has a positive real root of $x$ if
$\eta \in (0.96, 1]$ as $p\in[20,25]$, while if $\eta \in (0.97,
1]$ as $p\in[25,30]$. Below we present thoroughly studies on the
relations of several constraint equations and parameter space. For
simplicity, we take $\Lambda=0$ in (\ref{eq5}), which implies
${\cal A}_0=1$. From (\ref{eq29}) we can find a reasonable finite
algebraic root on $x$, but it is quite involved. We do not present
its expression here, instead we only express it formally
 \be
 x_{end}=x_{end}(p)
 \en
 From the scalar perturbation amplitude constraint Eq. (\ref{cobe})
 we get
 \be
 \label{eq39}
 x_i=\frac{2(-1+c+\eta)}{(-1+c)^2},
 \en
 where $c=y^{-1/3}(6\times 10^{-7}/p)^{1/3}$, and $y=\frac{V_i}{\mu
 ^4}$. Here the subscript $i$ means to take value at $k=aH$.
 Then the equation (\ref{eq34}) of the number of the e-folds can be
 reexpressed as
 \be
 \label{eq40}
 h(x_{end}(p))-h(x_i(p,y))=\frac{2N}{p}.
 \en
We note that there is a singularity in the function of $h(y,p)$,
 \be
 \lim \limits_{y\to y_s} h(y,p) \to -\infty ,
 \label{yinfi}
 \en
 where
 \be
 y_s=6\times 10^{-7}/p.
 \en
  This singularity appears when $x_i= \infty $ in (\ref{eq39}).
  The singularity Eq.~(\ref{yinfi})
 imposes a restriction on the energy scale of inflation: $y < y_s$.
 Otherwise, one may get a non-physical result: a negative number of
 the e-folds if one takes $y>y_s$. For a given $p$ and a required $N$, we
 always can get a $y$ (which satisfies $y<y_s$), so that the equation
 (\ref{eq40}) holds.

 In Fig.~1  we draw some contours of the number $N$ of e-folds
 through the relation (\ref{eq40}) with respect to $p$ and $x_i$.
 Note that here $x_i$ is determined by $y$ and $p$ through
 (\ref{eq39}).
 From the figure we can read the direct relation of $x_i$ and
$p$ for a given $N$.  For instance the curve just below the
notation ``N=60" gives the relation of $p$ to $x_i$ for the case
of $N=60$. In Fig.~2 we plot the relation among the parameters
$y$, $x_i$ and $p$ through (\ref{eq39}), which shows us the energy
scale of inflation. Note that in these figures $\eta =0.99$ is
taken. In this case, one has $\mu^2 = 0.01 \mu_p^2 \sim
(10^{17}Gev)^2$. So the energy scale of inflation in this setting
is $\rho_i \sim (10^{15}Gev)^4$, a reasonable inflation scale. In
Fig.~3 we plot the energy scale of inflation with respect to $N$
and $p$.

\begin{figure}
\centering
\includegraphics[totalheight=3in]{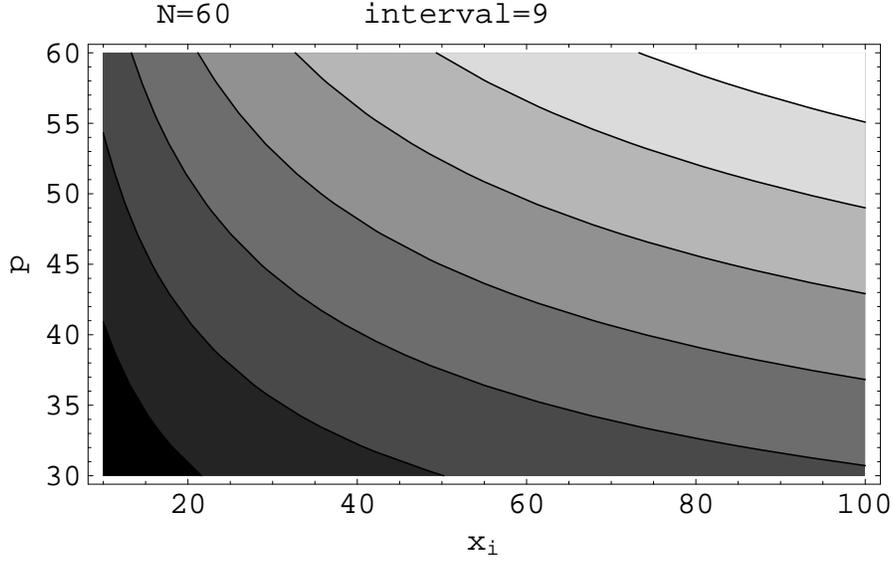}
\caption{The contours of the number $N$ of e-folds. The curve just
below the notation ``N=60" is the one for the case of $N=60$. The
curves on the right hand of that one represent cases with $N>60$,
and the interval is $\triangle N=9$.}
  \label{ncontour}
\end{figure}


\begin{figure}
\centering
\includegraphics[totalheight=3in]{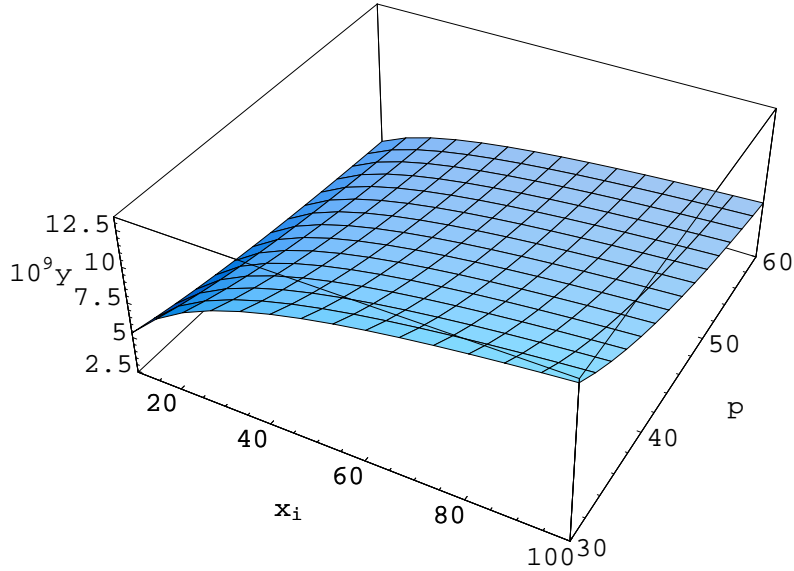}
\caption{The inflation energy scale $\frac{V_i}{\mu^4}$ versus the
parameters $x_i$ and $p$.} \label{y3d}
\end{figure}


\begin{figure}
\centering
\includegraphics[totalheight=3in]{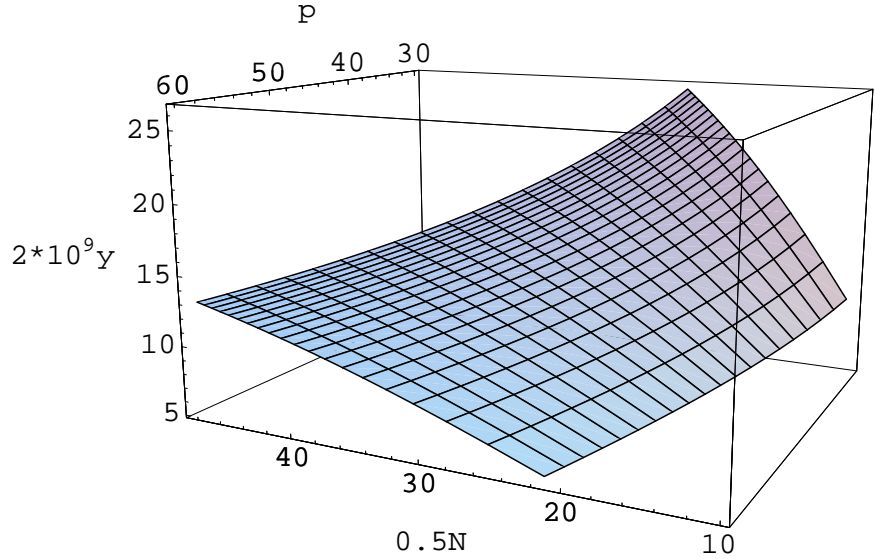}
\caption{The inflation energy scale $y$ versus the parameters $N$
and $p$}
 \label{npy}
\end{figure}


Now we consider a special case with fixed $N=60$ and $p=50$. In
this case, we have $x_{end}=0.05$, $x_i=36$ and $y= 2.4 \times
10^{-8}$. Furthermore we  find $\frac{\lambda}{\mu^4}=6.7\times
10^{-12}$, $\frac{m_5}{\mu}=0.02$,
$\frac{{}^{(5)}\Lambda}{\mu^2}=6.7\times10^{-14}$. In the Planck
unit $\mu_p\sim 10^{18}Gev$, they change to
$\frac{\lambda}{\mu_p^4}=6.7\times 10^{-16}$,
 $\frac{m_5}{\mu_p}=0.002$,
$\frac{{}^{(5)}\Lambda}{\mu_p^2}=6.7\times10^{-16}$.  For the
concrete model, in Figs. 4, 5, 6 we display the slow roll
parameters, scalar spectrum index and the running of scalar
spectrum index, respectively. Indeed, these quantities are in good
agreement with the observation data~\cite{wmap}.

\begin{figure}
\centering
\includegraphics[totalheight=2in]{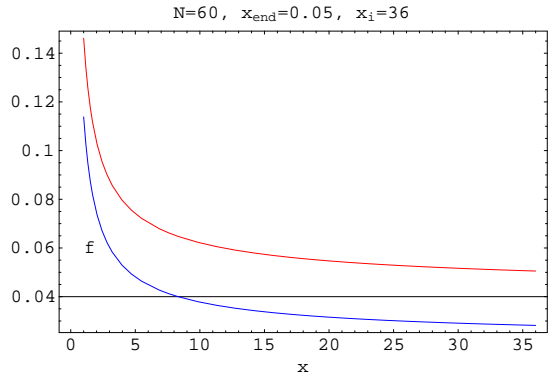}
\caption{The slow roll parameters $f$ and $\alpha$ versus $x$ for
the case of $N=60$ and $p=50$. The upper red curve represents
$\alpha$ and the bottom blue one stands for $f$.}
 \label{falpha}
\end{figure}
\begin{figure}
\centering
\includegraphics[totalheight=2in]{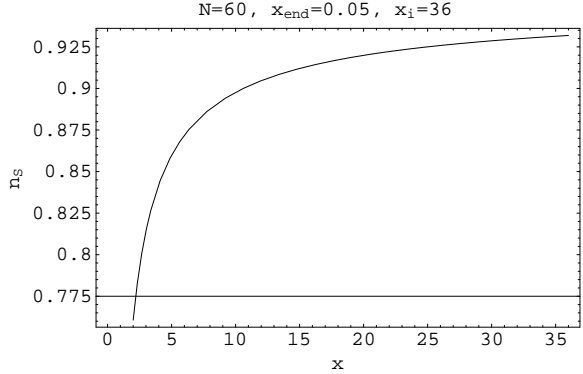}
\caption{The scalar spectrum index $n_S$ versus $x$ for the case
of $N=60$ and $p=50$.}
 \label{ns}
\end{figure}
\begin{figure}
\centering
\includegraphics[totalheight=2in]{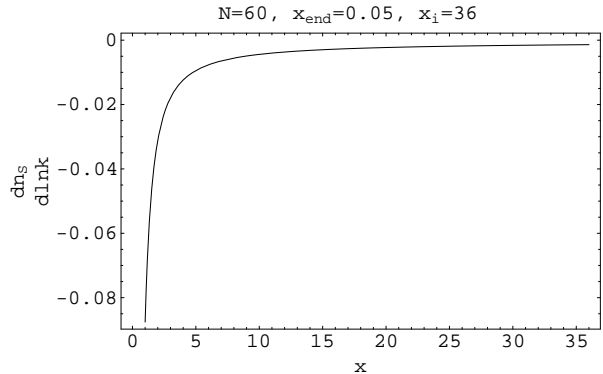}
\caption{The running of the scalar spectrum index
$\frac{dn_S}{d\ln k}$ versus $x$  for the case of $N=60$ and
$p=50$.}
 \label{run}
\end{figure}

\section{Conclusion and discussion}

Various inflationary models have been proposed since Guth's
seminal work~\cite{Guth}. Going with the observations of high
precision, we realize that a successful inflation model must at
least possess the following properties: 1) a sufficient large
number of e-folds, 2) a near Harrison-Zeldovich (scale invariant)
spectrum, 3) a negative running of spectrum index. The model we
discussed in this paper contents these. In particular, it is worth
noting that in this inflation model based on the warped DGP brane
world scenario, even for an exponential potential, the
inflationary phase can exit naturally, and within a reasonable
parameter region, the model can give us a negative running of the
scalar spectrum index. By choosing more appropriate parameters in
the model, we may obtain much better consistence with the current
observation data. These features are quite attractive. Other
properties of this model deserves further study. For example, it
would be interesting to investigate carefully the relations of the
spectrum index and its running to the comoving wave number $k$. In
addition, in the warped DGP brane model, it is certainly of
interest to further construct inflation models with scalar
spectrum index larger than one at larger scales.

{\bf Note added:} While we were finishing this paper, a
paper~\cite{new} appears on the net, which also discusses an
chaotic inflation model on a brane with induced gravity. More
recently, two related and interesting papers occur on the net. In
\cite{BMW} the authors calculate the amplitude of gravitational
waves from brane-world inflation with induced gravity, while the
authors of \cite{BW} study the induced gravity with a
non-minimally coupled scalar field on the brane.

{\bf Acknowledgments:} This work was supported in part by a grant
from Chinese academy of sciences, a grant No. 10325525 from NSFC,
and by the ministry of science and technology of China under grant
No. TG1999075401.

\end{document}